\documentclass[aps,prd,onecolumn,preprintnumbers,showpacs,nofootinbib,amssymb]{revtex4}
\usepackage{graphicx}
\usepackage{amsmath}
\usepackage{amssymb}
\usepackage{amsfonts}
\usepackage{bm}

\def\be{\begin{equation}}
\def\ee{\end{equation}}
\def\bea{\begin{eqnarray}}
\def\eea{\end{eqnarray}}

\def\nn{\nonumber \\}
\def\e{\mathrm{e}}

\begin{document}

\title{$f\left(R,T\right)$ gravity}

\author{Tiberiu Harko}
\email{harko@hkucc.hku.hk}
\affiliation{Department of Physics and
Center for Theoretical and Computational Physics, The University
of Hong Kong, Pok Fu Lam Road, Hong Kong, P. R. China}

\author{Francisco S. N. Lobo}
\email{flobo@cii.fc.ul.pt}
\affiliation{Centro de Astronomia e Astrof\'{\i}sica da
Universidade de Lisboa, Campo Grande, Ed. C8 1749-016 Lisboa,
Portugal}

\author{Shin'ichi Nojiri}
\email{nojiri@phys.nagoya-u.ac.jp}
\affiliation{Department of Physics, Nagoya University, Nagoya 464-8602,
Japan}
\affiliation{Kobayashi-Maskawa Institute for the Origin of Particles and the
Universe,
Nagoya University, Nagoya 464-8602, Japan}

\author{Sergei D. Odintsov}
\email{odintsov@aliga.ieec.uab.es}
\affiliation{Institucio Catalana de Recerca i Estudis Avancats
(ICREA) and Institut de Ciencies de lEspai (IEEC-CSIC),
Campus UAB, Facultat de Ciencies, Torre C5-Par-2a pl,
E-08193 Bellaterra (Barcelona), Spain \\
Also at Tomsk State Pedagogical University, Tomsk }

\begin{abstract}
We consider $f(R,T)$ modified theories of gravity, where the
gravitational Lagrangian is given by an arbitrary function of the
Ricci scalar $R$ and of the trace of the stress-energy tensor $T$.
We obtain the gravitational field equations in the metric
formalism, as well as the equations of motion for test particles,
which follow from the covariant divergence of the stress-energy
tensor. Generally, the gravitational field equations depend on the
nature of the matter source. The field equations of several
particular models, corresponding to some explicit forms of the
function $f(R,T)$, are also presented. An important case,
which is analyzed in detail, is represented by scalar field
models. We write down the action and briefly consider the
cosmological implications of the $f\left(R,T^{\phi}\right)$
models, where $T^{\phi}$ is the trace of the stress-energy tensor
of a self-interacting scalar field. The equations of motion of the
test particles are also obtained from a variational principle. The
motion of massive test particles is non-geodesic, and takes place
in the presence of an extra force orthogonal to the four-velocity.
The Newtonian limit of the equation of motion is further analyzed.
Finally, we provide a constraint on the magnitude of the
extra-acceleration by analyzing the perihelion precession of the
planet Mercury in the framework of the present model.
\end{abstract}


\date{\today}

\maketitle

\section{Introduction}

The recent observational data \cite{Ri98,PeRa03} on the late-time
acceleration of the Universe and the existence of dark matter have
posed a fundamental theoretical challenge to gravitational
theories. One possibility in explaining the observations is by
assuming that at large scales the Einstein gravity model of
general relativity breaks down, and a more general action
describes the gravitational field. Theoretical models, in which
the standard Einstein-Hilbert action is replaced by an arbitrary
function of the Ricci scalar $R$ \cite{RevNoOd}, have been
extensively investigated lately. The presence of a late-time
cosmic acceleration of the Universe can indeed be explained by
$f(R)$ gravity \cite{Carroll:2003wy}. The conditions of the
existence of viable cosmological models have been found in
\cite{viablemodels}, and severe weak field constraints obtained
from the classical tests of general relativity for the Solar
System regime seem to rule out most of the models proposed so far
\cite{solartests,Olmo07}. However, viable models, passing Solar
System tests, can be constructed
\cite{Hu:2007nk,solartests2,Sawicki:2007tf,Amendola:2007nt}.
$f(R)$ models that satisfy local tests and unify inflation with
dark energy were considered in \cite{odin}. In the framework of
$f(R)$ gravity models the possibility that the galactic dynamic of
massive test particles can be understood without the need for dark
matter was considered in
\cite{Cap2,Borowiec:2006qr,Mar1,Boehmer:2007kx,Bohmer:2007fh}. For
reviews of $f(R)$ generalized gravity models see \cite{RevNoOd,
SoFa08}.

A generalization of $f(R)$ modified theories of gravity was
proposed in \cite{Bertolami:2007gv}, by including in the theory an
explicit coupling of an arbitrary function of the Ricci scalar $R$
with the matter Lagrangian density $L_\mathrm{m}$. As a result of
the coupling the motion of the massive particles is non-geodesic,
and an extra force, orthogonal to the four-velocity, arises. The
connections with Modified Newtonian Dynamics (MOND) and the
Pioneer anomaly were also explored. This model was extended to the
case of the arbitrary couplings in both geometry and matter in
\cite{ha08}. The astrophysical and cosmological implications of
the non-minimal coupling  matter-geometry coupling were
extensively investigated in \cite{Bertolami:2007vu, ha10}, and the
Palatini formulation of the non-minimal geometry-coupling models
was considered in \cite{Pal}. In this context, a maximal extension
of the Hilbert-Einstein action was proposed in \cite{Rlm}, by
assuming that the gravitational Lagrangian is given by an
arbitrary function of the Ricci scalar $R$ and of the matter
Lagrangian $L_\mathrm{m}$. The gravitational field equations have
been obtained in the metric formalism, as well as the equations of
motion for test particles, which follow from the covariant
divergence of the stress-energy tensor.

A specific application of the latter $f(R,L_\mathrm{m})$ gravity
was proposed in \cite{Poplawski:2006ey}, which may be considered a
relativistically covariant model of interacting dark energy, based
on the principle of least action. The cosmological constant in the
gravitational Lagrangian is a function of the trace of the
stress-energy tensor, and consequently the model was denoted
``$\Lambda(T)$ gravity''. It was argued that recent cosmological
data favor a variable cosmological constant, which are consistent
with $\Lambda(T)$ gravity, without the need to specify an exact
form of the function $\Lambda(T)$ \cite{Poplawski:2006ey}.
$\Lambda(T)$ gravity is more general than the Palatini $f(R)$
gravity, and reduces to the latter when we neglect the pressure of
the matter.

It is the purpose of the present paper to consider another
extension of standard general relativity, the $f(R,T)$ modified
theories of gravity, where the gravitational Lagrangian is given
by an arbitrary function of the Ricci scalar $R$ and of the trace
of the stress-energy tensor $T$. Note that the dependence from $T$
may be induced by exotic imperfect fluids or quantum effects
(conformal anomaly). As a first step in our study we derive the
field equations of the model from a variational, Hilbert-Einstein
type, principle. The covariant divergence of the stress-energy
tensor is also obtained. The $f(R,T)$ gravity model depends on a
source term, representing the variation of the matter
stress-energy tensor with respect to the metric. A general
expression for this source term is obtained as a function of the
matter Lagrangian $L_\mathrm{m}$. Therefore each choice of
$L_\mathrm{m}$ would generate a specific set of field equations.
Some particular models, corresponding to specific choices of the
function $f(R,T)$ are also presented, and their properties are
briefly discussed. In fact, we also demonstrate the possibility of
reconstruction of arbitrary FRW cosmologies by an appropriate
choice of a function $f(T)$. Scalar fields play a fundamental role
in cosmology, i.e., as possible explanations for inflation, late
time acceleration, or dark matter, respectively. Therefore, we
introduce and briefly discuss the $f\left(R,T^{\phi }\right)$
gravitational models, where $T^{\phi }$ is the trace of the
stress-energy of the scalar field. Some cosmological applications
of this model are also presented.

Since in the present model the covariant divergence of the
stress-energy tensor is non-zero, the motion of massive test
particles is non-geodesic, and an extra acceleration, due to the
coupling between matter and geometry, is always present. The
equations of motion of test particles are obtained from a
variational principle. The same variational principle can be used
to investigate the Newtonian limit of the model, and the
expression of the extra-acceleration is also obtained. We use the
precession of the perihelion of the planet Mercury to obtain a
general constraint on the magnitude of the extra-acceleration.

The present paper is organized as follows. The field equations of
$f\left(R,T\right)$ gravity are derived in Section~\ref{1}. Some
particular cases of the model are considered in
Section~\ref{part}. The case of the scalar fields is discussed in
Section~\ref{Od}, and a Brans-Dicke type formulation of the model
is obtained. The equations of motion of massive test particles are
derived in Section~\ref{3}, where the Newtonian limit of the model
is also obtained and analyzed. We discuss and conclude our results
in Section~\ref{disc}. In the present paper we use the natural
system of units with $G=c=1$, so that the Einstein gravitational
constant is defined as $\kappa^2 =8\pi$.

\section{Gravitational field equations of $f\left(R,T\right)$
gravity}\label{1}

We assume that the action for the modified theories of gravity  takes the
following form
\begin{equation}
S=\frac{1}{16\pi}\int
f\left(R,T\right)\sqrt{-g}\;d^{4}x+\int{L_\mathrm{m}\sqrt{-g}\;d^{4}x}\, ,
\end{equation}
where $f\left(R,T\right)$ is an arbitrary function of the Ricci
scalar, $R$, and of the trace $T$ of the stress-energy tensor of
the matter, $T_{\mu \nu}$. $L_\mathrm{m}$ is the matter Lagrangian
density, and we define the stress-energy tensor of matter as
\cite{LaLi}
\begin{equation}\label{enmom}
T_{\mu \nu }=-\frac{2}{\sqrt{-g}}
\frac{\delta \left( \sqrt{-g}L_\mathrm{m}\right) }{\delta g^{\mu \nu }}\, ,
\end{equation}
and its trace by $T=g^{\mu \nu }T_{\mu \nu }$, respectively. By
assuming that the Lagrangian density $L_\mathrm{m}$ of matter
depends only on the metric tensor components $g_{\mu \nu }$, and
not on its derivatives, we obtain
\begin{equation}\label{en1}
T_{\mu \nu }=g_{\mu \nu }L_\mathrm{m}-2
\frac{\partial L_\mathrm{m}}{\partial g^{\mu\nu}}\, .
\end{equation}

By varying the action $S$ of the gravitational field with respect
to the metric tensor components $g^{\mu \nu }$ provides the
following relationship
\begin{eqnarray}
\delta S&=&\frac{1}{16\pi }\int \left[ f_{R}\left( R,T\right)
\delta R+f_{T}\left(R,T\right) \frac{\delta T}{\delta g^{\mu \nu }}
\delta g^{\mu \nu }-\frac{1}{2}g_{\mu \nu }f\left( R,T\right) \delta g^{\mu
\nu }
+16\pi \frac{1}{\sqrt{-g}}
\frac{\delta \left( \sqrt{-g}L_\mathrm{m}\right) }{\delta g^{\mu \nu
}}\right]
\sqrt{-g}d^{4}x\, ,
\end{eqnarray}
where we have denoted  $f_{R}\left( R,T\right) =\partial
f\left(R,T\right) /\partial R$ and $f_T\left( R,T\right) =\partial
f\left(R,T\right) /\partial T$, respectively. For the variation of
the Ricci scalar, we obtain
\begin{equation}
\delta R=\delta \left( g^{\mu \nu }R_{\mu \nu }\right) =R_{\mu \nu }\delta
g^{\mu \nu }+g^{\mu \nu }\left( \nabla _{\lambda }
\delta \Gamma _{\mu \nu}^{\lambda }-\nabla _{\nu }\delta \Gamma _{\mu
\lambda }^{\lambda }\right)\, ,
\end{equation}
where $\nabla_{\lambda }$ is the covariant derivative with respect
to the symmetric connection $\Gamma $ associated to the metric
$g$. The variation of the Christoffel symbols yields
\begin{equation}
\delta \Gamma _{\mu \nu }^{\lambda }=\frac{1}{2}g^{\lambda \alpha }\left(
\nabla _{\mu }\delta g_{\nu \alpha }+\nabla _{\nu }
\delta g_{\alpha \mu}-\nabla _{\alpha }\delta g_{\mu \nu }\right)\, ,
\end{equation}
and the variation of the Ricci scalar provides the expression
\begin{equation}
\delta R=R_{\mu \nu }\delta g^{\mu \nu }+g_{\mu \nu }\square
\delta g^{\mu\nu }-\nabla _{\mu }\nabla _{\nu }\delta g^{\mu \nu }\, .
\end{equation}

Therefore, for the variation of the action of the gravitational
field we obtain
\begin{eqnarray}
\delta S&=&\frac{1}{16\pi }\int \Big[ f_{R}\left( R,T\right) R_{\mu \nu }
\delta g^{\mu\nu }+f_{R}\left( R,T\right) g_{\mu \nu }\square
\delta g^{\mu \nu}-f_{R}\left( R,T\right) \nabla _{\mu }\nabla _{\nu }\delta
g^{\mu \nu}
\nonumber  \\
&& +f_{T}\left( R,T\right) \frac{\delta \left(g^{\alpha \beta }T_{\alpha
\beta }\right)}{\delta
g^{\mu \nu }} \delta g^{\mu \nu }-\frac{1}{2}g_{\mu \nu }f\left( R,T\right)
\delta
g^{\mu \nu }+16\pi \frac{1}{\sqrt{-g}}
\frac{\delta \left( \sqrt{-g}L_\mathrm{m}\right) }{\delta g^{\mu \nu
}}\Bigg]
\sqrt{-g}d^{4}x\, .  \label{var1}
\end{eqnarray}
We define the variation of $T$ with respect to the metric tensor as
\begin{equation}
\frac{\delta \left(g^{\alpha \beta }T_{\alpha \beta }\right)}{\delta g^{\mu
\nu}}
=T_{\mu\nu}+\Theta _{\mu \nu}\, ,
\end{equation}
where
\begin{equation}
\Theta_{\mu \nu}\equiv g^{\alpha \beta }\frac{\delta T_{\alpha \beta
}}{\delta g^{\mu \nu}}\, .
\end{equation}
After partially integrating the second and third terms in Eq.~(\ref{var1}),
we obtain the field equations of the $f\left( R,T\right) $ gravity model as
\begin{eqnarray}\label{field}
f_{R}\left( R,T\right) R_{\mu \nu } - \frac{1}{2}
f\left( R,T\right)  g_{\mu \nu }
+\left( g_{\mu \nu }\square -\nabla_{\mu }\nabla _{\nu }\right)
f_{R}\left( R,T\right) =8\pi T_{\mu \nu}-f_{T}\left( R,T\right)
T_{\mu \nu }-f_T\left( R,T\right)\Theta _{\mu \nu}\, .
\end{eqnarray}
Note that when $f(R,T)\equiv f(R)$, from Eqs.~(\ref{field}) we
obtain the field equations of $f(R)$ gravity.

By contracting Eq.~(\ref{field}) gives the following relation
between the Ricci scalar $R$ and the trace $T$ of the
stress-energy tensor,
\begin{eqnarray}
&&f_{R}\left( R,T\right) R+3\square f_{R}\left( R,T\right) -2
f\left( R,T\right) =8\pi T-f_{T}\left(R,T\right)
T-f_{T}\left(R,T\right)\Theta\, ,
\label{contr}
\end{eqnarray}
where we have denoted $\Theta =\Theta^{\ \mu}_{\mu}$.

By eliminating the term $\square f_{R}\left( R,T\right)$ between
Eqs.~(\ref{field}) and (\ref{contr}), the gravitational field
equations can be written in the form
\begin{eqnarray}
f_{R}\left( R,T\right) \left( R_{\mu \nu }-\frac{1}{3}Rg_{\mu \nu
}\right) +\frac{1}{6} f\left( R,T\right)  g_{\mu\nu }
&=&8\pi \left(T_{\mu \nu}-\frac{1}{3}T
g_{\mu \nu}\right)-f_{T}\left( R,T\right)
\left( T_{\mu \nu }-\frac{1}{3}T g_{\mu \nu }\right)
\nonumber  \\
&&-f_T\left(R,T\right)\left(\Theta_{\mu \nu}-\frac{1}{3}
\Theta g_{\mu\nu}\right)
+ \nabla _{\mu }\nabla _{\nu }f_{R}\left(R,T\right)\, .
\end{eqnarray}
Taking into account the covariant divergence of Eq.~(\ref{field}),
with the use of the following mathematical identity \cite{Ko06}
\begin{eqnarray}
\nabla ^{\mu }\left[ f_R\left(R,T\right)
R_{\mu\nu}-\frac{1}{2}f\left(R,T\right)g_{\mu\nu}
+\left(g_{\mu \nu }\square -\nabla_{\mu }\nabla_{\nu}\right)
f_R\left(R,T\right)\right] \equiv 0\, ,
\end{eqnarray}
where $f\left(R,T\right)$ is an arbitrary function of the Ricci
scalar $R$ and of the trace of the stress-energy tensor $T$, we
obtain for the divergence of the stress-energy tensor $T_{\mu
\nu}$ the equation
\begin{equation}\label{noncons}
\nabla ^{\mu }T_{\mu \nu }
=\frac{f_{T}\left( R,T\right) }{8\pi -f_{T}\left(R,T\right) }
\left[ \left( T_{\mu \nu }+\Theta _{\mu \nu }\right) \nabla^{\mu }
\ln f_{T}\left( R,T\right) +\nabla ^{\mu }\Theta _{\mu \nu }\right]\, .
\end{equation}

Next we consider the calculation of the tensor $\Theta _{\mu
\nu}$, once the matter Lagrangian is known. From Eq.~(\ref{en1})
we obtain first
\begin{eqnarray}
\frac{\delta T_{\alpha \beta }}{\delta g^{\mu \nu}}
&=&\frac{\delta g_{\alpha \beta }}{\delta g^{\mu\nu}}L_\mathrm{m}
+g_{\alpha \beta }\frac{\partial L_\mathrm{m}}{\partial g^{\mu
\nu}}-2\frac{\partial ^2L_\mathrm{m}}
{\partial g^{\mu \nu }\partial g^{\alpha \beta }} \nonumber\\
&=&\frac{\delta g_{\alpha \beta }}{\delta g^{\mu \nu }}L_\mathrm{m}
+\frac{1}{2}g_{\alpha \beta }g_{\mu\nu}L_\mathrm{m}-\frac{1}{2}g_{\alpha
\beta }
T_{\mu \nu }-2\frac{\partial ^2L_\mathrm{m}}{\partial g^{\mu \nu }\partial
g^{\alpha \beta }}\, .
\end{eqnarray}
From the condition $g_{\alpha \sigma }g^{\sigma \beta }=\delta
_{\alpha }^{\beta }$, we have
\begin{equation}
\frac{\delta g_{\alpha \beta }}{\delta g^{\mu \nu }}
=-g_{\alpha \sigma }g_{\beta \gamma }\delta ^{\sigma \gamma }_{\mu \nu }\, ,
\end{equation}
where $\delta ^{\sigma \gamma }_{\mu \nu }=\delta g^{\sigma \gamma
}/\delta g^{\mu \nu}$ is the generalized Kronecker symbol.
Therefore for $\Theta_{\mu \nu}$ we find
\begin{equation}\label{var}
\Theta _{\mu \nu}=-2T_{\mu \nu}+g_{\mu \nu }L_\mathrm{m}-2g^{\alpha \beta }
\frac{\partial ^2L_\mathrm{m}}{\partial g^{\mu \nu }\partial g^{\alpha \beta
}}\, .
\end{equation}
In the case of the electromagnetic field the matter Lagrangian is
given by
\begin{equation}
L_\mathrm{m}=-\frac{1}{16\pi }F_{\alpha \beta }F_{\gamma \sigma }
g^{\alpha \gamma }g^{\beta \sigma }\, ,
\end{equation}
where $F_{\alpha \beta }$ is the electromagnetic field tensor. In
this case we obtain $\Theta _{\mu\nu}=-T_{\mu \nu }$. In the case
of a massless scalar field $\phi $ with Lagrangian
$L_\mathrm{m}=g^{\alpha \beta }\nabla_{\alpha }\phi \nabla _{\beta
}\phi $, we obtain $\Theta _{\mu \nu}=-T_{\mu \nu}+(1/2)Tg_{\mu
\nu}$. The problem of the perfect fluids, described by an energy
density $\rho $, pressure $p$ and four-velocity $u^{\mu}$ is more
complicated, since there is no unique definition of the matter
Lagrangian. However, in the present study we {\it assume} that the
stress-energy tensor of the matter is given by
\begin{equation}\label{entens}
T_{\mu \nu}=\left(\rho +p\right)u_{\mu }u_{\nu}-pg_{\mu \nu}\, ,
\end{equation}
and the matter Lagrangian can be taken as $L_\mathrm{m}=-p$. The
four-velocity $u_{\mu }$ satisfies the conditions
$u_{\mu}u^{\mu}=1$ and $u^{\mu }\nabla _{\nu }u_{\mu }=0$,
respectively. Then, with the use of Eq.~(\ref{var}), we obtain for
the variation of the stress-energy of a perfect fluid the
expression
\begin{equation}
\Theta _{\mu \nu }=-2T_{\mu \nu }-pg_{\mu \nu }\, .
\end{equation}

\section{Particular cases of gravitational field equations in the $f(R,T)$
model}\label{part}

In the present Section we consider some particular classes of
$f(R,T)$ modified gravity models, obtained by explicitly
specifying the functional form of $f$. Generally, the field
equations also depend, through the tensor $\Theta _{\mu \nu }$, on
the physical nature of the matter field. Hence in the case of
$f(R,T)$ gravity, depending on the nature of the matter source,
for each choice of $f$ we can obtain several theoretical models,
corresponding to different matter models.

\subsection{$f\left(R,T\right)=R+2f(T)$}

As a first case of a $f(R,T)$ modified gravity model we assume
that the function $f(R,T)$ is given by
$f\left(R,T\right)=R+2f(T)$, where $f(T)$ is an arbitrary function
of the trace of the stress-energy tensor of matter. The
gravitational field equations immediately follow from
Eq.~(\ref{field}), and are given by
\begin{equation}
R_{\mu\nu}-\frac{1}{2}Rg_{\mu\nu}
=8\pi T_{\mu \nu }-2f'\left(T\right)T_{\mu\nu}-2f'(T)
\Theta _{\mu \nu}+f(T)g_{\mu \nu }\, ,
\end{equation}
where the prime denotes a derivative with respect to the argument.

If the matter source is a perfect fluid, $\Theta _{\mu
\nu}=-2T_{\mu\nu}-pg_{\mu \nu}$, then the field equations become
\begin{equation}
R_{\mu\nu}-\frac{1}{2}Rg_{\mu\nu}=8\pi T_{\mu\nu}
+2f'\left(T\right)T_{\mu\nu}+\left[2pf'(T)+f(T)\right]g_{\mu \nu }\, .
\end{equation}
In the case of dust with $p=0$ the gravitational field equations
are given by
\begin{equation}
R_{\mu\nu}-\frac{1}{2}Rg_{\mu\nu}
=8\pi T_{\mu\nu}+2f'(T)T_{\mu\nu}+f(T)g_{\mu\nu}\, .
\label{Ein1}
\end{equation}
These field equations were proposed in \cite{Poplawski:2006ey} to
solve the cosmological constant problem. The simplest cosmological
model can be obtained by assuming a dust universe ($p=0$, $T=\rho
$), and by choosing the function $f(T)$ so that $f(T)=\lambda T$,
where $\lambda $ is a constant. By assuming that the metric of the
universe is given by the flat Robertson-Walker metric, \be
ds^2=dt^2-a^2(t)\left(dx^2+dy^2+dz^2\right)\, , \ee the
gravitational field equations are given by
\begin{eqnarray}
3\frac{\dot{a}^2}{a^2}&=&\left(8\pi +3\lambda \right)\rho\, , \\
2\frac{\ddot{a}}{a}+\frac{\dot{a}^2}{a^2}&=&\lambda \rho\, ,
\end{eqnarray}
respectively. Thus this $f(R,T)$ gravity model is equivalent to a
cosmological model with an effective cosmological constant
$\Lambda _\mathrm{eff}\propto H^2$, where $H=\dot{a}/a$ is the
Hubble function \cite{Poplawski:2006ey}. It is also interesting to
note that generally for this choice of $f(R,T)$ the gravitational
coupling becomes an effective and time dependent coupling, of the
form $G_\mathrm{eff}=G\pm 2f'(T)$. Thus the term $2f(T)$ in the
gravitational action modifies the gravitational interaction
between matter and curvature, replacing $G$ by a running
gravitational coupling parameter.

The field equations reduce to a single equation for $H$,
\be
2\dot{H}+3\frac{8\pi +2\lambda }{8\pi +3\lambda }H^2=0\, ,
\ee
with the general solution given by
\be H(t)=\frac{2\left(8\pi
+3\lambda \right)}{3\left(8\pi +2\lambda \right)}\frac{1}{t}\, .
\ee
The scale factor evolves according to $a(t)=t^{\alpha }$, with
$\alpha ={2\left(8\pi +3\lambda \right)/3\left(8\pi +2\lambda
\right)}$.

\subsection{$f\left(R,T\right)=f_1(R)+f_2(T)$}

As a second example we consider the case in which the function $f$
is given by $f\left(R,T\right)=f_1(R)+f_2(T)$, where $f_1(R)$ and
$f_2(T)$ are arbitrary functions of $R$ and $T$, respectively. In
this case for an arbitrary matter source the gravitational field
equations are given by
\begin{eqnarray}
f'_1(R)R_{\mu\nu}-\frac{1}{2}f_1(R)g_{\mu\nu}+(g_{\mu\nu} \square-\nabla_\mu
\nabla_\nu)f_1(R)
%\nonumber\\
=8\pi T_{\mu\nu}-f_2'(T)T_{\mu\nu}-f_2'(T)\Theta_{\mu\nu}
+\frac{1}{2}f_2T)g_{\mu\nu}\, .
\label{Ein2}
\end{eqnarray}
Assuming that the matter content consists of a perfect fluid, the
gravitational field equations become
\begin{eqnarray}
f'_1(R)R_{\mu\nu}-\frac{1}{2}f_1(R)g_{\mu\nu}+(g_{\mu\nu} \square-\nabla_\mu
\nabla_\nu)f'_1(R)=
%\nonumber\\
8\pi
T_{\mu\nu}+f_2'(T)T_{\mu\nu}+\left[f_2'(T)p+\frac{1}{2}f_2(T)\right]g_{\mu\nu}\,
.
\label{Ein21}
\end{eqnarray}
In the case of dust with $p=0$, the gravitational field equations
reduce to
\begin{eqnarray}
f'_1(R)R_{\mu\nu}-\frac{1}{2}f_1(R)g_{\mu\nu}+(g_{\mu\nu} \square-\nabla_\mu
\nabla_\nu)f'_1(R)=
%\nonumber\\
8\pi T_{\mu\nu}+f_2'(T)T_{\mu\nu}+\frac{1}{2}f_2(T)g_{\mu\nu}\, .
\label{Ein22}
\end{eqnarray}
In the case $f_2(T)\equiv 0$, we re-obtain the field equations of
standard $f(R)$ gravity. Eq.~(\ref{Ein21}) can be reformulated as
an effective Einstein field equations of the form
\be G_{\mu
\nu}=R_{\mu \nu }-\frac{1}{2}Rg_{\mu \nu} =8\pi
G_\mathrm{eff}T_{\mu\nu} +T_{\mu \nu }^\mathrm{eff}\, ,
\ee
where
we have denoted
\be G_\mathrm{eff}=\frac{1}{f_1^{\prime }(R)}
\left[1+\frac{f_2^{\prime }(T)}{8\pi }\right]\, ,
\ee
and
\be
T_{\mu \nu }^\mathrm{eff}=\frac{1}{f_1^{\prime }(R)}
\left\{\frac{1}{2}\left[f_1(R)-Rf_1^{\prime }(R) +2f^{\prime
}_2(T)p+f_2(T)\right]g_{\mu \nu}-\left(g_{\mu\nu}
\square-\nabla_\mu \nabla_\nu \right)f'_1(R)\right\}\, .
\ee
The gravitational coupling is again given by an effective, matter
(and time) dependent coupling, which is proportional to the
derivative of the function $f_2$ with respect to $T$. The
gravitational field equations can be recast in such a form that
the higher order corrections, coming both from the geometry, and
from the matter-geometry coupling, provide a stress-energy tensor
of geometrical and matter origin, describing an ``effective''
source term on the right hand side of the standard Einstein field
equations. In the $f(R,T)$ scenario, the cosmic acceleration may
result not only from  a geometrical contribution to the total
cosmic energy density, but it is also dependent on the matter
content of the universe, which provides new corrections to the
Hilbert-Einstein Lagrangian via the matter-geometry coupling.

The $(t,t)$ component of Eq.~(\ref{Ein22}) has the following form:
\be
\label{tt1}
3H^2 = \frac{8\pi}{f_1^{\prime}(R)}\left[ 1+\frac{f_2^{\prime }(T)}{8\pi
}\right] \rho +\frac{1}{f_1^{\prime}(R)}\left[-\frac{1}{2}\left(f_1(R)
-6 \left(\dot H + 2 H^2 \right) f_1^\prime(R) \right)+2f_2^\prime (T)
-9 \left(\ddot H + 4 H\dot H\right) f_1^{\prime\prime}(R)\right] \, .
\ee
Here $R=6 \left( \dot H + 2 H^2 \right)$. For simplicity, $T_{\mu\nu}$
correspondsto the matter with a constant EoS parameter $w$.
If we now define the $e$-folding $N$ by $a=a_0 \e^N$, $\rho$ and $T$ are
given by
\be
\label{tt2}
\rho = \rho_0 \e^{-3(1+w)N}\, ,\quad T = - (1-3w) \rho_0 \e^{-3(1+w)N}\, .
\ee

We now consider an arbitrary development of the expansion in the
Universe given by
\be \label{tt3}
H = h(N)\, ,
\ee
where $h(N)$ is an arbitrary function of $N$. Then Eq.~(\ref{tt1})
can be written as
\bea \label{tt4} f_2^\prime (T) &=& F_2 (N) \nn & \equiv &
\frac{3}{1+ \rho_0 \e^{-3(1+w)N}} \left\{ - \frac{8\pi}{3} \rho_0
\e^{-3(1+w)N} + \frac{1}{6} f_1 \left[ 6 \left( h(N) h^\prime(N) +
2 h(N)^2 \right) \right] \right. \nn && - \left[h(N) h^\prime(N) +
h(N)^2 \right] f_1^\prime \left[ 6 \left( h(N) h^\prime(N) + 2
h(N)^2 \right) \right] \nn && \left. + 3 \left[ h(N)^2
h^{\prime\prime} (N) + h(N) h'(N)^2 + 4 h(N)^2 h'(N) \right]
f_1^{\prime\prime} \left[ 6 \left( h(N) h^\prime(N) + 2 h(N)^2
\right) \right] \right\} \, . \eea
Eq.~(\ref{tt4}) dictates that for an arbitrary $f_1(R)$, and for
the following specific choice
\be \label{tt5}
f_2^\prime (T) = F_2
\left( - \frac{\ln \left( -
\frac{T}{\left(1-3w\right)\rho_0}\right)}{3(1+w)} \right) \, ,
\ee
an arbitrary development of the expansion in the Universe given by
(\ref{tt3}) can be realized. Hence, for viable $f(R)$
gravitational models, using the above reconstruction method, the
possibility arises to modify the universe evolution by adding the
corresponding function depending on the trace of the stress-energy
tensor.

\subsection{$f\left(R,T\right)=f_1(R)+f_2(R)f_3(T)$}

As a third case of generalized $f(R,T)$ gravity models, we
consider that the action is given by
$f\left(R,T\right)=f_1(R)+f_2(R)f_3(T)$, where $f_i$, $i=1,2,3$
are arbitrary functions of the argument. For an arbitrary matter
source the gravitational field equations are given by
\begin{eqnarray}
\left[f'_1(R)+f'_2(R)f_3(T)\right]R_{\mu\nu}-\frac{1}{2}f_1(R)g_{\mu\nu}
+(g_{\mu\nu} \square-\nabla_\mu
\nabla_\nu)\left[f'_1(R)+f'_2(R)f_3(T)\right]
\nonumber   \\
=8\pi T_{\mu\nu}-f_2(R)f_3'(T)T_{\mu\nu}-f_2(R)f_3'(T)\Theta _{\mu\nu}
+\frac{1}{2}f_2(R)f_3(T)g_{\mu\nu}\, .
\label{Ein3}
\end{eqnarray}
In the case of a perfect fluid we find the field equations
\begin{eqnarray}
\left[f'_1(R)+f'_2(R)f_3(T)\right]R_{\mu\nu}-\frac{1}{2}f_1(R)g_{\mu\nu}
+(g_{\mu\nu} \square-\nabla_\mu
\nabla_\nu)\left[f'_1(R)+f'_2(R)f_3(T)\right]
\nonumber \\
=8\pi T_{\mu\nu}+f_2(R)f_3'(T)T_{\mu\nu}+f_2(R)\left[f_3'(T)p
+\frac{1}{2}f_3(T)\right]g_{\mu \nu}\, .
\label{Ein31}
\end{eqnarray}
For the case of dust matter we obtain
\begin{eqnarray}
\left[f'_1(R)+f'_2(R)f_3(T)\right]R_{\mu\nu}-\frac{1}{2}f_1(R)g_{\mu\nu}
+(g_{\mu\nu} \square-\nabla_\mu
\nabla_\nu)\left[f'_1(R)+f'_2(R)f_3(T)\right]
\nonumber   \\
=8\pi T_{\mu\nu}+f_2(R)f_3'(T)T_{\mu\nu}+\frac{1}{2}f_2(R)f_3(T)g_{\mu
\nu}\, .
\label{Ein32}
\end{eqnarray}
In this class of models both, the effective cosmological constant
$\Lambda _\mathrm{eff}$ and the running gravitational coupling
$G_\mathrm{eff}$ are functions of both matter and geometry.

\section{$f\left(R,T^{\phi}\right)$ gravity}\label{Od}

Scalar fields are supposed to play a fundamental role in physics
and cosmology \cite{Faraoni}. In particular, cosmological
inflation, the late-time acceleration of the universe, or dark
matter and its properties can be explained in the framework of
specific scalar field models. However, obtaining more general
gravitational models with scalar fields as a source may give a
better insight in the general properties of the gravitational
field, and could also provide some possibilities for
observationally testing the generalizations of gravity models. In
the present Section, we consider the $f(R,T)$ gravity model in the
case of self-interacting scalar fields.

\subsection{The action of the $f\left(R,T^{\phi}\right)$ gravity}

We start with the following action for matter,
\be
S_\mathrm{matter} \left( g_{\mu\nu},\psi_i \right)
= \int d^4x \sqrt{-g} \mathcal{L} \left( g_{\rho\sigma}, \psi_i \right)\, .
\label{FRT0}
\ee
In Eq.~(\ref{FRT0}) the $\psi_i$'s, $i=1,2,...$ represent the
matter fields. By using the matter action (\ref{FRT0}), we now
introduce the action for the gravitational field with matter
sources as
\be \label{FRT1}
S =\frac{1}{2\kappa^2} \int d^4x
\sqrt{-g} F(R,\phi) + S_\mathrm{matter} \left( \e^\phi
g_{\mu\nu},\psi_i \right)\, ,
\ee
where $F(R,\phi)$ is an algebraic function of $R$ and of the
scalar field $\phi$. Then by the variation of the action with
respect to $\phi$, we obtain first
\be \label{FRT2}
\frac{1}{2\kappa^2}F_\phi (R,\phi) + \frac{1}{2}T_\phi =0\, ,
\ee
where we denoted $F_\phi (R,\phi) \equiv \partial F (,\phi) /
\partial \phi$ and
\be \label{FRT3}
T_\phi \equiv \e^{-\phi}
g^{\mu\nu} T_{\phi\,\mu\nu}, T_{\phi\,\mu\nu} \equiv \left.
T_{\mu\nu} \right|_{g_{\mu\nu} \to \e^\phi g_{\mu\nu} }\, ,
\ee
respectively. The stress-energy tensor of matter is defined, as
usual, by
\begin{equation}
\quad T_{\mu\nu} \equiv -\frac{2}{\sqrt{-g}}
\frac{\delta S_\mathrm{matter} \left( g_{\rho\sigma},\psi_i \right)}
{\delta g_{\mu\nu}}\, .
\end{equation}
By the assumption that $F(R,\phi)$ is an algebraic function of $R$ and
$\phi$,
Eq.~(\ref{FRT2}) can be algebraically solved with respect to $\phi$.
Thus we can obtain $\phi $ as
a function of $R$ and $T_\phi$, i.e., $\phi = \phi \left( R, T_\phi
\right)$.
Then by substituting the expression of $\phi $ into the action (\ref{FRT1}),
we obtain an example of $F(R,T_\phi)$ gravity, with the following action
\be
\label{FRT4}
S = \frac{1}{2\kappa^2}\int d^4x \sqrt{-g} \tilde F(R,T_\phi)
+ S_\mathrm{matter} \left( \e^\phi g_{\mu\nu},\psi_i \right)\, ,
\ee
where we have denoted
\begin{equation}
\quad
\tilde F\left(R,T_\phi\right) \equiv F \left[ R, \phi\left(R,T_\phi\right)
\right]\, .
\end{equation}
With the use of the conformal Weyl transformation $g_{\mu\nu} \to \e^{-\phi}
g_{\mu\nu}$, the action
(\ref{FRT1}) or (\ref{FRT4}) is transformed as
\bea
\label{FRT5}
S &=& \frac{1}{2\kappa^2}\int d^4 x \sqrt{-g} \e^{-2\phi} F\left[ \left(R +
3 \Box \phi
 - \frac{3}{2} \partial_\sigma \phi \partial^\sigma \phi \right)
\e^\phi ,\phi\right]
+ S_\mathrm{matter} \left( g_{\mu\nu},\psi_i \right) \nn
&& = \frac{1}{2\kappa^2}\int d^4x \sqrt{-g} \tilde F\left[ \left(R
+ 3 \Box \phi
 - \frac{3}{2} \partial_\sigma \phi \partial^\sigma \phi \right)\e^\phi,
T\right]
+ S_\mathrm{matter} \left( g_{\mu\nu},\psi_i \right)\ ,
\eea
with $T\equiv g^{\mu\nu} T_{\mu\nu}$. In the action
$S_\mathrm{matter} \left( g_{\mu\nu},\psi_i \right)$ in
Eq.~(\ref{FRT5}), the matter fields have only a minimal coupling
with gravity, and they do not couple with $\phi$. Then the frame
in the action (\ref{FRT5}) might be regarded as a physical frame.

\subsection{Example of $f\left(R,T^{\phi}\right)$ scalar field gravity, and
reconstruction}

As an example of $f\left(R,T^{\phi}\right)$ gravity of the form
$R+f\left(T^{\phi }\right)$, we consider the case of a scalar
field with a self-interaction potential $V\left(\phi \right)$. The
action is given by
\be \label{Phi1} S^\phi = \int d^4 x \sqrt{-g}
\left[ - \frac{1}{2}\omega(\phi)
\partial_\mu \phi \partial^\mu \phi - V(\phi) \right]\, ,
\ee
where we have included $\omega(\phi)$ for later convenience.
For the scalar field model described by Eq.~(\ref{Phi1}),
the trace of the stress-energy tensor is given by
\be
\label{Phi2}
T^\phi = - \omega(\phi)
\partial_\mu \phi \partial^\mu \phi - 4V(\phi)\, .
\ee
Consequently we may define the $f\left(R,T^{\phi}\right)=R+f\left(T^{\phi
}\right)$
gravity model in the following form:
\be
\label{Phi3}
S = \int d^4 x \sqrt{-g} \left[\frac{1}{2\kappa^2}R + f (T^\phi) -
\frac{1}{2}\omega(\phi)
\partial_\mu \phi \partial^\mu \phi - V(\phi) \right]\, .
\ee
For the model (\ref{Phi3}), in a flat Friedman-Robertson-Walker geometry,
the Friedman equations have the following form:
\begin{eqnarray}
\frac{3}{\kappa^2} H^2 &=& \frac{1}{2} \omega(\phi) {\dot\phi}^2 + V (\phi)
  - f\left[ \omega(\phi) {\dot\phi}^2 - 4 V (\phi) \right]
+ 2 f'\left[ \omega(\phi) {\dot\phi}^2 - 4 V (\phi) \right]
\omega(\phi) {\dot\phi}^2\, ,\label{Phi4}\\
- \frac{1}{\kappa^2} \left(3 H^2 + 2 \dot H \right) &=& \frac{1}{2}
\omega(\phi) {\dot\phi}^2 - V (\phi)
+ f\left[ \omega(\phi) {\dot\phi}^2 - 4 V (\phi) \right]\, ,\label{Phi5}
\end{eqnarray}
where $H=\dot{a}/a$.

In the following, we consider, for simplicity,  the case $V(\phi)=0$.
Then the action (\ref{Phi3}) has the following form:
\be
\label{Phi6}
S = \int d^4 x \sqrt{-g} \left\{ \frac{1}{2\kappa^2}R + F \left[ -
\omega(\phi)
\partial_\mu \phi \partial^\mu \phi\right] \right\}\, ,
\ee
and the Friedman equations (\ref{Phi4}) and (\ref{Phi5}) take the form,
\bea
\label{Phi7}
\frac{3}{\kappa^2} H^2 &=& - F\left[ \omega(\phi) {\dot\phi}^2 \right]
+ 2 F'\left[ \omega(\phi) {\dot\phi}^2 \right] \omega(\phi) {\dot\phi}^2\,
,\\
\label{Phi8}
  - \frac{1}{\kappa^2} \left(3 H^2 + 2 \dot H \right)
&=& F\left[ \omega(\phi) {\dot\phi}^2 \right]\, .
\eea
In the action (\ref{Phi6}), $F \left[ - \omega(\phi)
\partial_\mu \phi \partial^\mu \phi\right] $ is defined by
\be
\label{Phi9}
F \left[ - \omega(\phi)
\partial_\mu \phi \partial^\mu \phi\right] \equiv f \left[ - \omega(\phi)
\partial_\mu \phi \partial^\mu \phi\right] - \frac{1}{2}\omega(\phi)
\partial_\mu \phi \partial^\mu \phi\, .
\ee
The action (\ref{Phi6}) gives a model of k-essence
\cite{Chiba:1999ka, ArmendarizPicon:2000dh, ArmendarizPicon:2000ah}.
In \cite{Matsumoto:2010uv}, it has been shown that the Friedmann equations
(\ref{Phi7}) and
(\ref{Phi8}) do not admit the de Sitter solution, except in the trivial case
where $\phi$ is a constant, and $F(0)>0$.
In \cite{Matsumoto:2010uv}, the formalism of the general reconstruction has
also been explicitly
given. An explicit model of modified gravity in which a crossing of the
phantom divide can be
realized was reconstructed in \cite{odin1}.

As a simple example, we consider the model
\be
\label{Phi10}
F \left[ - \omega(\phi)
\partial_\mu \phi \partial^\mu \phi\right] = - F_0
\e^{- 2 \ln \left(\frac{\phi}{\phi_0}\right) \partial_\mu \phi \partial^\mu
\phi }\, ,
\ee
where $F_0$ and $\phi_0$ are constants. The Friedman equations have a
solution where the universe
expands by a power law,
\be
\label{Phi11}
H = \frac{h_0}{t}\, ,\quad \phi=t.
\ee
The constant $h_0$ can be obtained  by solving the following algebraic
equation
\be
\label{Phi12}
3h_0^2 - 2h_0 = \kappa^2 \phi_0^2 F_0 \, .
\ee

\section{The equation of motion of test particles and the Newtonian limit in
$f\left(R,T\right)$ gravity}\label{3}

Since in the general $f(R,T)$ type gravity models the
stress-energy tensor of matter is not covariantly conserved, it
follows that the test particles, moving in a gravitational field,
do not follow geodesic lines. This situation is similar to the
case of the $f\left(R,L_\mathrm{m}\right)$ models \cite{Rlm},
where the coupling between matter and geometry induces a
supplementary acceleration acting on the particle. In the present
Section, we derive the equations of motion of test particles in
$f(R,T)$ gravity models, and obtain the Newtonian limit of the
theory. We also investigate the constraints on the magnitude of
the extra-acceleration that can be obtained from the available
observational data on the perihelion precession of the planet
Mercury.

\subsection{The equations of motion of test particles}

In the case of a perfect fluid, with the stress-energy tensor given by
Eq.~(\ref{entens}), the
divergence of the stress-energy tensor is given by
\begin{equation}\label{cons2}
\nabla ^{\mu }T_{\mu \nu }=-\frac{1}{8\pi +f_{T}\left( R,T\right) }\left\{
T_{\mu \nu }\nabla ^{\mu }f_{T}\left( R,T\right) +g_{\mu \nu }\nabla ^{\mu }
\left[ f_{T}\left( R,T\right) p\right] \right\}\, .
\end{equation}
We also introduce the projection operator
$h_{\mu \lambda }=g_{\mu\lambda}-u_{\mu }u_{\lambda }$
for which we have $h_{\mu \lambda }u^{\mu}=0$ and
$h_{\mu \lambda }T^{\mu\nu}=-h_{\lambda }^{\nu}p$, respectively.

Explicitly, Eq.~(\ref{cons2}) can be written in the form
\begin{eqnarray}\label{precontr}
&&\nabla _{\nu }\left( \rho +p\right)u^{\mu }u^{\nu }+\left(
\rho +p\right) \left[ u^{\nu }\nabla _{\nu }u^{\mu }
+u^{\mu }\nabla _{\nu}u^{\nu}\right] -g^{\mu \nu }\nabla _{\nu }p
\nonumber\\
&&=-\frac{1}{8\pi +f_{T}\left( R,T\right) }\left\{
T^{\mu \nu }\nabla _{\nu }f_{T}\left( R,T\right)
+g^{\mu \nu }\nabla _{\nu}
\left[ f_{T}\left( R,T\right) p\right] \right\}\, .
\end{eqnarray}
By contracting Eq.~(\ref{precontr}) with $h_{\mu \lambda }$ we obtain
\begin{equation}
g_{\mu \lambda }u^{\nu }\nabla _{\nu }u^{\mu }
=8\pi\frac{\nabla _{\nu }p}{\left(\rho +p\right)\left[8\pi
+f_{T}\left(R,T\right)\right]} h_{\lambda }^{\nu }\, .
\end{equation}
After multiplying with $g^{\alpha \lambda }$ and by taking into account the
identity
\begin{equation}
u^{\nu }\nabla _{\nu }u^{\mu }=\frac{d^{2}x^{\mu }}{ds^{2}}
+\Gamma _{\nu\lambda}^{\mu}u^{\nu }u^{\lambda }\, ,
\end{equation}
we obtain the equation of motion of a test fluid in $f\left(R,T\right)$
gravity as
\begin{equation}
\frac{d^{2}x^{\mu }}{ds^{2}}+\Gamma _{\nu \lambda }^{\mu }u^{\nu }u^{\lambda
}=f^{\mu }\, ,  \label{eqmot}
\end{equation}
where
\begin{equation}
f^{\mu }=8\pi\frac{\nabla _{\nu }p}{\left(\rho +p\right)\left[8\pi
+f_{T}\left(R,T\right)\right]}\left(g^{\mu \nu }-u^{\mu }u^{\nu }\right)\, .
\end{equation}
The extra-force $f^{\mu }$  is perpendicular to the four-velocity,
$f^{\mu}u_{\mu }=0$.
When $f_T\left(R,T\right)=0$, we re-obtain the equation of motion of perfect
fluids
with pressure in standard general relativity, which follow from the conservation of the energy-momentum tensor, $\nabla _{\mu }T_{\nu }^{\mu }=0$ \cite{Sp}. In the limit $p\rightarrow 0$, corresponding to a pressureless fluid (dust), in standard general relativity the motion of the test particles becomes geodesic. The same result holds true in the $f(R,T)$ gravity model. Even if $f_T\left(R,T\right)\neq 0$, the motion of the dust particles always follows the geodesic lines of the geometry.
By assuming that the term $8\pi\nabla_{\nu }p/\left(\rho +p\right)\left[8\pi
+f_{T}\left(R,T\right)\right]$
can be formally represented as $\nabla _{\nu }\ln \sqrt{Q}$,
\begin{equation}\label{Q}
8\pi\frac{\nabla _{\nu }p}{\left(\rho +p\right)\left[8\pi
+f_{T}\left(R,T\right)\right]}=\nabla _{\nu }\ln \sqrt{Q}\, ,
\end{equation}
the equation of motion Eq.~(\ref{eqmot}) can be obtained from the
variational principle
\begin{equation}
\delta S_{p}=\delta \int L_{p}ds=\delta \int \sqrt{Q}\sqrt{g_{\mu \nu
}u^{\mu }u^{\nu }}ds=0\, ,  \label{actpart}
\end{equation}
where $S_{p}$ and $L_{p}=\sqrt{Q}\sqrt{g_{\mu \nu }u^{\mu }u^{\nu
}}$ are the action and the Lagrangian density for the test
particles, respectively.

To prove this result we start with the Lagrange equations
corresponding to the action~(\ref{actpart}),
\begin{equation}
\frac{d}{ds}\left( \frac{\partial L_{p}}{\partial u^{\lambda }}\right)
 - \frac{\partial L_{p}}{\partial x^{\lambda }}=0\, .
\end{equation}
Since
\begin{equation}
\frac{\partial L_{p}}{\partial u^{\lambda }}=\sqrt{Q}u_{\lambda }
\end{equation}
and
\begin{equation}
\frac{\partial L_{p}}{\partial x^{\lambda }}
=\frac{1}{2} \sqrt{Q}g_{\mu \nu,\lambda }u^{\mu }u^{\nu }
+\frac{ 1}{2} \frac{Q_{,\lambda }}{Q}\, ,
\end{equation}
a straightforward calculation gives the equations of motion of the
particle as
\begin{equation}
\frac{d^{2}x^{\mu }}{ds^{2}}+\Gamma _{\nu \lambda }^{\mu }u^{\nu }u^{\lambda
}+\left( u^{\mu }u^{\nu }-g^{\mu \nu }\right) \nabla _{\nu }\ln \sqrt{Q}=0
\, .
\end{equation}
When $\sqrt{Q}\rightarrow 1$ we re-obtain the standard general
relativistic equation for geodesic motion.

As an example of the application of the previous formalism we
consider the case in which the pressure can be expressed as a
function of the density by a linear barotropic equation of state
of the form $p=w\rho $, where the constant $w$ satisfies the
condition $w\ll 1$. Therefore $\rho +p\approx \rho $ and $T=\rho
-3p\approx \rho$, respectively. Moreover, for simplicity, we also
assume that the function $f_T$ is a function of $T\approx \rho $
only. We can expand $f_T$ near a fixed value $\rho_0$ of the
density, so that $f_T\left(\rho \right)=f_T\left(\rho _0\right)
+\left(\rho -\rho _0\right)\left[df_T\left/d\rho \right)/d\rho
\right]|_{\rho=\rho_0} =8\pi \left[a_0+ b_0\left(\rho -\rho
_0\right)\right]$, where $a_0=f_T\left(\rho _0\right)/8\pi $ and
$b_0=\left[df_T\left/d\rho \right)/d\rho
\right]|_{\rho=\rho_0}/8\pi$, respectively. Eq.~(\ref{Q}) of the
definition of $\sqrt{Q}$ becomes
\begin{equation}\label{Q1}
\frac{w}{1+a_0-b_0\rho _0}\nabla _{\nu }\ln\frac{\rho }{1+a_0+b_0\left(\rho
-\rho _0\right)}
=\nabla _{\nu }\ln \sqrt{Q}\, ,
\end{equation}
giving
\begin{equation}\label{Q2}
\sqrt{Q}\left(\rho \right)\approx
\left[\frac{C\rho }{1+a_0+b_0\left(\rho -\rho
_0\right)}\right]^{w/\left(1+a_0-b_0\rho _0\right)}\, ,
\end{equation}
where $C$ is an arbitrary constant of integration. Eq.~(\ref{Q}) is also valid for a fluid satisfying a linear barotropic equation of state of the form $p=\left(\gamma -1\right)\rho $, $\gamma ={\rm constant}$, and for a model with $f_T(R,T)={\rm constant}=f_T$. In this case $\sqrt{Q}=C_T\rho ^{8\pi (\gamma -1)/\gamma \left(8\pi +f_T\right)}$, where $C_T$ is an arbitrary integration constant. Therefore Eq.~(\ref{Q}) is valid in both the non-relativistic and the extreme relativistic limits of the model. On the other hand we have to mention that the function $\sqrt{Q}$ can always be obtained by formally integrating the left-hand side of Eq.~(\ref{Q}). However, generally this function cannot be expressed in an exact analytical form, and to find its functional form approximate methods have to be used. 

\subsection{The Newtonian limit}

The variational principle~(\ref{actpart}) and the pressureless
dust model, described by Eqs.~(\ref{Q1}) and (\ref{Q2}), can be
used to study the Newtonian limit of the model. In the limit of
the weak gravitational fields,
\begin{equation}
ds\approx \sqrt{1+2\phi -\vec{v}^{2}}dt
\approx \left( 1+\phi -\vec{v}^{2}/2\right) dt\, ,
\end{equation}
where $\phi $ is the Newtonian potential and $\vec{v}$ is
the usual tridimensional velocity of the fluid.
By using the relation $x^{\alpha }=\exp (\alpha \ln x)\approx 1+\alpha
\ln x$,
we can approximate $\sqrt{Q}\left(\rho \right)$ given by Eq.~(\ref{Q2}) as
\be
\sqrt{Q}\left(\rho \right)\approx 1+\frac{w}{\left(1+a_0-b_0\rho _0\right)}
\ln \left[\frac{C\rho }{1+a_0+b_0\left(\rho -\rho
_0\right)}\right]=1+U\left(\rho \right)\, ,
\ee
where we have denoted
\be
U\left(\rho \right)=\frac{w}{\left(1+a_0-b_0\rho _0\right)}
\ln \left[\frac{C\rho }{1+a_0+b_0\left(\rho -\rho _0\right)}\right]\, .
\ee

In the first order of approximation the equations of motion of the
fluid can be derived from the variational principle
\begin{equation}
\delta \int \left[ 1+U\left(\rho \right)
+\phi -\frac{\vec{v}^{2}}{2}\right] dt=0\, ,
\end{equation}
and are given by
\begin{equation}
\vec{a}=-\nabla \phi -\nabla U\left(\rho
\right)=\vec{a}_{N}+\vec{a}_p+\vec{a}_{E}\, ,
\end{equation}
where $\vec{a}$ is the total acceleration of the system,
$\vec{a}_{N}=-\nabla \phi $ is the Newtonian gravitational
acceleration and
\be\label{fidr} \vec{a}_p
=-\frac{C}{1+a_0-b_0\rho _0}\frac{1}{\rho }\nabla p=-\frac{1}{\rho
} \nabla p , \ee
is the hydrodynamical acceleration. Eq.~(\ref{fidr}) also allows
us to fix the value of the arbitrary integration constant $C$ as
$C=1+a_0-b_0\rho _0$. Finally,
\begin{equation}
\vec{a}_{E}\left(\rho ,p\right)
=\frac{b_0}{1+a_0-b_0\rho _0}\frac{\nabla p}{1+a_0+b_0\left(\rho -\rho
_0\right)}\, ,
\end{equation}
is a supplementary acceleration induced due to the modification of the
action of the gravitational field.

\subsection{The precession of the perihelion of Mercury}

An estimation of the effect of the extra-force, generated by the
coupling between matter and geometry, on the orbital parameters of
the motion of the planets around the Sun can be obtained in a
simple way by using the properties of the Runge-Lenz vector,
defined as $\vec{A}=\vec{v}$ $\times \vec{L}-\alpha \vec{e}_{r}$,
where $\vec{v}$ is the velocity relative to the Sun, with mass
$M_{\odot}$, of a planet of mass $m$, $\vec{r}=r\vec{e}_{r}$ is
the two-body position vector, $\vec{p}=\mu \vec{v}$ is the
relative momentum,  $\mu =mM_{\odot }/\left( m+M_{\odot}\right) $
is the reduced mass, $\ \ \vec{L}=\vec{r}$ $\times \vec{p}=\mu
r^{2}\dot{\theta}\vec{k}$ is the angular momentum, and $\alpha
=GmM_{\odot}$ \cite{prec}. For an elliptical orbit of eccentricity
$e$, major semi-axis $a$, and period $T$, the equation of the
orbit is given by $\left( L^{2}/\mu\alpha \right) r^{-1}=1+e\cos
\theta $. The Runge-Lenz vector can be expressed as
\begin{equation}
\vec{A}=\left( \frac{\vec{L}^{2}}{\mu r}-\alpha \right)
\vec{e}_{r}-\dot{r}L\vec{e}_{\theta }\, ,
\end{equation}
and its  derivative with respect to the polar
angle $\theta $ is given by
\begin{equation}
\frac{d\vec{A}}{d\theta }
=r^{2}\left[ \frac{dV(r)}{dr}-\frac{\alpha}{r^{2}}\right] \vec{e}_{\theta
}\, ,
\end{equation}
where $V(r)$ is the potential of the central force \cite{prec}.
The potential term consists of the Post-Newtonian potential,
$V_{PN}(r)=-\alpha /r-3\alpha ^{2}/mr^{2}$, plus the contribution
from the general coupling between matter and geometry. Thus we
have
\begin{equation}
\frac{d\vec{A}}{d\theta }=r^{2}\left[ 6\frac{\alpha ^{2}}{mr^{3}}
+m\vec{a}_{E}(r)\right] \vec{e}_{\theta }\, ,
\end{equation}
where we have also assumed that $\mu \approx m$. The change in
direction $\Delta \phi $ of the perihelion with a change of
$\theta $ of $2\pi $ is obtained as $\Delta\phi =\left( 1/\alpha
e\right) \int_{0}^{2\pi }\left\vert \dot{\vec{L}}\times
d\vec{A}/d\theta \right\vert d\theta $, and it is given by
\begin{equation}\label{prec}
\Delta \phi =24\pi ^{3}\left( \frac{a}{T}\right) ^{2}\frac{1}{1-\e^{2}}
+\frac{L}{8\pi ^{3}me}
\frac{\left( 1-e^{2}\right) ^{3/2}}{\left( a/T\right)^{3}}
\int_{0}^{2\pi }\frac{a_{E}\left[ L^{2}
\left( 1+\e\cos \theta \right)^{-1}/m\alpha \right] }{\left( 1+e\cos \theta
\right)^{2}}
\cos \theta d\theta\, ,
\end{equation}%
where we have used the relation $\alpha /L=2\pi \left( a/T\right)
/\sqrt{1-e^{2}}$. The first term of this equation corresponds to
the standard general relativistic precession of the perihelion of
the planets, while the second term gives the contribution to the
perihelion precession due to the presence of the coupling between
matter and geometry.

As an example of the application  of Eq.~(\ref{prec}) we consider
the case for which the extra-force may be considered as a
constant,  $a_E\approx$ constant, an approximation that could be
valid for small regions of spacetime. In the Newtonian limit the
extra-acceleration generated by the coupling between matter and
geometry can be expressed in a similar form
\cite{Bertolami:2007gv}. With the use of Eq.~(\ref{prec}) one
finds for the perihelion precession the expression
\begin{equation}\label{prec1}
\Delta \phi =\frac{6\pi GM_{\odot}}{a\left( 1-e^{2}\right) }+\frac{2\pi
a^{2}
\sqrt{1-e^{2}}}{GM_{\odot}}a_{E}\, ,
\end{equation}
where we have also used Kepler's third law, $T^2=4\pi
^2a^3/GM_{\odot}$. For the planet Mercury $a=57.91\times 10^{11}$
cm, and $e=0.205615$, respectively, while $M_{\odot }=1.989\times
10^{33}$ g. With these numerical values the first term in Eq.
(\ref{prec1}) gives the standard general relativistic value for
the precession angle, $\left( \Delta \phi \right) _{GR}=42.962$
arcsec per century, while the observed value of the precession is
$\left(\Delta \phi \right)_{obs}=43.11\pm0.21$ arcsec per century
\cite{merc}. Therefore the difference $\left(\Delta \phi
\right)_{E}=\left(\Delta \phi \right)_{obs}-\left( \Delta \phi
\right) _{GR}=0.17$ arcsec per century can be attributed to other
physical effects. Hence the observational constraints requires
that the value of the constant $a_E$ must satisfy the condition
$a_E\leq 1.28\times 10^{-9}$ cm/s$^2$.

\section{Discussions and final remarks}\label{disc}

In the present paper we have considered a generalized gravity
model with an arbitrary coupling between matter (described by the
trace of the stress-energy tensor) and geometry, with the
Lagrangian given by an arbitrary function of $T$ and  of the Ricci
scalar. We have derived the gravitational field equations
corresponding to this model, and considered several particular
cases that may be relevant in explaining some of the open problems
of cosmology and astrophysics. The new
matter and time dependent terms in the gravitational field
equations play the role of an effective cosmological
constant. We have also demonstrated the possibility of
reconstruction of arbitrary FRW cosmologies by an appropriate
choice of a function $f(T)$. The equations of motion corresponding
to this model show the presence of an extra-force acting on test
particles, and the motion is generally non-geodesic. We have
obtained, by using the perihelion precession of Mercury, an upper
limit on the magnitude of the extra-acceleration in the Solar
System. This value of $a_E$, obtained from the solar system
observations, is somewhat smaller than the value of the
extra-acceleration $a_{E}\approx 10^{-8}$ cm/s$^{2}$, necessary to
explain the ``dark matter'' properties, as well as the Pioneer
anomaly \cite{Bertolami:2007gv, BPT06, BPL07}. However, it does
not rule out the possibility of the presence of some extra
gravitational effects acting at both the solar system and galactic
levels, since the assumption of a constant extra-force may not be
correct on larger astronomical scales.

Therefore the predictions of the $f(R,T)$ gravity model could lead
to some major differences, as compared to the predictions of
standard general relativity, or other generalized gravity models,
in several problems of current interest, such as cosmology,
gravitational collapse or the generation of gravitational waves.
The study of these phenomena may also provide some specific
signatures and effects, which could distinguish and discriminate
between the various gravitational models. In order to explore in
more detail the connections between the $f(R,T)$ gravity model and
the cosmological evolution, some explicit physical models are
necessary to be built. This will be done in forthcoming work.

\section*{Acknowledgments}

The work of TH was supported by an GRF grant of the government of
the Hong Kong SAR. FSNL acknowledges financial support of the
Funda\c{c}\~{a}o para a Ci\^{e}ncia e Tecnologia through the
grants PTDC/FIS/102742/2008, CERN/FP/109381/2009 and
CERN/FP/116398/2010. This research was also supported in part by
MEC (Spain) project FIS2006-02842 and AGAUR(Catalonia) 2009SGR-994
(SDO), by Global COE Program of Nagoya University (G07) provided
by the Ministry of Education, Culture, Sports, Science \&
Technology and by the JSPS Grant-in-Aid for Scientific Research
(S) \# 22224003 (SN).

\end{document}